\documentclass[pra,aps,floatfix,twocolumn]{revtex4}
\usepackage{graphicx}
\usepackage{bm}

\begin{document}

\title{Magneto optical trapping of Barium}

\author{S. De, U. Dammalapati\footnote{Present address: University of Strathclyde, Glasgow, UK},
K. Jungmann and L.
Willmann}
\address{Kernfysisch Versneller Instituut, University of Groningen,
Zernikelaan 25, 9747 AA Groningen, The Netherlands}

\begin{abstract}
First laser cooling and trapping of the heavy alkaline earth element
barium has been achieved based on the strong
6s$^2$~$^1$S$_0$~-~6s6p~$^1$P$_1$ transition for the main cooling.
Due to the large branching into metastable D-states several
additional laser driven transitions are required to provide a closed
cooling cycle. A total efficiency of $0.4(1) \cdot 10^{-2}$ for
slowing a thermal atomic beam and capturing atoms into a magneto
optical trap was obtained. Trapping lifetimes of more than 1.5~s
were observed. This lifetime is shortened at high laser intensities
by photo ionization losses. The developed techniques will allow to
extend significantly the number of elements that can be optically
cooled and trapped.
\end{abstract}

\maketitle

The heavy alkaline earth element barium (Ba) has been laser cooled
and captured in a magneto-optical trap (MOT). Of particular interest
is the trap loading efficiency from a thermal atomic beam because
the developed technique is essential to trap short lived isotopes of
the chemical homologue radium for searches for permanent electric
dipole moments (EDM's) \cite{flambaum1999} or measurements of atomic
parity violation (APV) \cite{flambaum2000}.

Laser cooling and trapping of neutral atoms relies on narrow band
optical excitation in an almost closed subset of atomic states. In
general, this requires driving of more than one optical transition.
Laser cooled atoms have become a vital tool for a variety of
experiments, e.g. Bose-Einstein condensation, high precision
measurements, optical frequency standards and studies on fundamental
symmetries. To date, the list of optically trapped elements includes
all alkaline metals \cite{LIN1991}, noble gases in metastable states
\cite{ASPECT1988} except Rn, alkaline earth elements (Mg
\cite{SENGSTOCK1994}, Ca \cite{calciumtrap}, Sr \cite{calciumtrap}
and Ra \cite{GUEST2007}) and several other elements, e.g. Cr
\cite{BELL1999}, Er \cite{MCCLELLAND2006}, Ag \cite{UHLENBERG2000}
Yb \cite{HONDA1999}, Hg \cite{hachisu2007} and Cd
\cite{brinkman2007}. Extending laser cooling to other elements
requires the selection of an efficient laser cooling scheme which
accounts for peculiarities of their atomic level structure.

\begin{figure}[b]
\center
\includegraphics[width = 65mm, angle = 0]{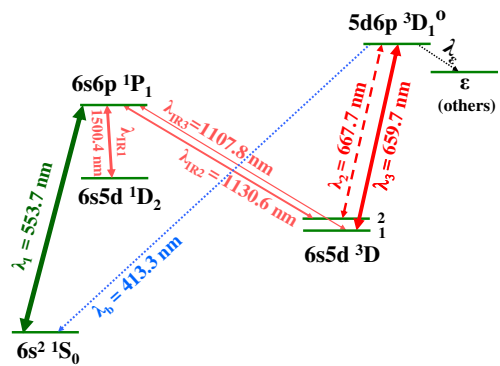}\\
\caption{Low lying energy levels of atomic Ba relevant for laser
cooling. Full lines indicate laser driven transitions and dashed
lines show spontaneous decay channels.} \label{balev}
\end{figure}

In Ba or Ra the strong ns$^2$~$^{1}$S$_{0}$-nsnp~$^{1}$P$_{1}$
transitions, n=6, 7, offer large optical forces. However, in both
cases the substantial branching fractions of 0.3$\%$ of the
nsnp~$^{1}$P$_{1}$ states to metastable D-states require
quantitative repumping via several transitions
\cite{DAMMALAPATI2006} (Fig.\ref{balev}). Only in Ra, the
7s$^2$~$^{1}$S$_{0}$-7s7p~$^{3}$P$_{1}$ intercombination line offers
an alternative for cooling. Although the losses from the
$^3$P$_1$-state are less than 4$\cdot 10^{-5}$, its rather long
lifetime of 422(20)ns \cite{scielzo2006} results in a cooling force
which is smaller by two orders of magnitude. The advantages of this
cooling scheme are the lower Doppler cooling limit and the
simplicity of repumping, but for slowing the efficiency  is rather
limited. Capturing of Ra in a MOT was reported from a Zeeman slowed
atomic beam with 7$\cdot$10$^{-7}$ efficiency \cite{GUEST2007}.

\begin{figure*}[bt]
\center
\includegraphics[width = 76 mm, angle = 270]{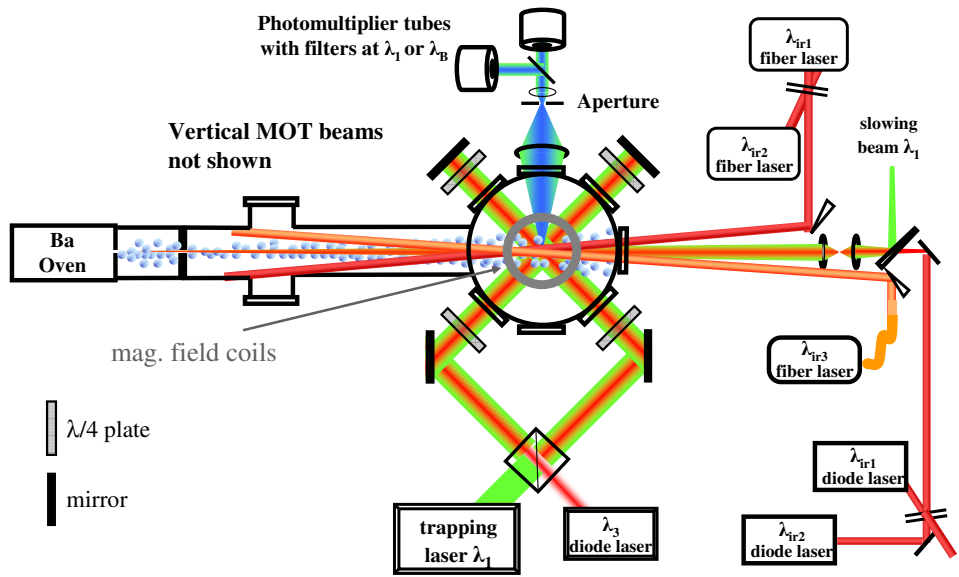}\
\ \caption{Setup for laser cooling and trapping of barium.}
\label{coolsetup}
\end{figure*}

The ns$^2$~$^1$S$_0$-nsnp~$^1$P$_1$ transitions, n=3...5, are used
for laser cooling and trapping of lighter alkaline earth elements,
where the branching to metastable D-states is much smaller than for
Ra and Ba. On average, an atom is transferred to one of the
metastable D-states after scattering of only
$\rm{A_{leak}^{Ba}}$=330(30) photons at wavelength $\lambda_1$ for
Ba (Fig. \ref{balev} and Tab. \ref{bawavelengths}). This corresponds
to a velocity change of 1.8(2) m/s only. The largest leak from a
cooling cycle of previously trapped elements of
$\rm{A_{leak}^{Cr}}$=~2500 was given for Cr~\cite{BELL1999}. Because
of its lighter mass Cr could be loaded into a magneto optical trap
without repumping, however, the efficiency increased by two orders
of magnitude with repumping from the D-states. In contrast, for Ba
no trapping can be expected without effective repumping from all
three low-lying D-states, i.e. 6s5d~$^1$D$_2$, 6s5d~$^3$D$_1$ and
6s5d~$^3$D$_2$ (Fig. \ref{balev}).
\begin{table}[tb]
\caption{Vacuum wavelengths and experimental transition rates for
barium.} \label{bawavelengths}
\begin{tabular}{|c|c|c|r|l|}
\hline
Upper  &  Lower & ~Label~ & ~Wavelength~ & ~~A$_{ik}$ \\
Level   & Level& & [nm]~~ & ~[10$^{8}$]~s~$^{-1}$ \\
\hline \hline
6s6p~$^{1}$P$_{1}$~ & 6s$^2$~$^{1}$S$_{0}$ & $\lambda_1$ & 553.7~ & ~~1.19(1)$^{\rm a}$ \\
& ~6s5d~$^{1}$D$_{2}~$ & $\lambda_{\rm{ir1}}$  & 1500.4~ & ~~0.0025(2)$^{\rm a}$ \\
 & ~6s5d~$^{3}$D$_{2}~$ & $\lambda_{\rm{ir2}}$ & 1130.6~ & ~~0.0011(2)$^{\rm a}$ \\
& ~6s5d~$^{3}$D$_{1}~$ & $\lambda_{\rm{ir3}}$  & 1107.8~ & ~~0.000031(5)$^{\rm a}$ \\
\hline
~5d6p~$^3$D$_1^\circ$~ & 6s$^2$~$^{1}$S$_{0}$ & $\lambda_{\rm B}$ & 413.3~ & ~~0.013(1)$^{\rm b}$ \\
 & 6s5d~$^{3}$D$_{2}$ & $\lambda_{2}$ & 667.7~ & ~~0.17(2)$^{\rm a}$\\
 & 6s5d~$^{3}$D$_{1}$ & $\lambda_{3}$ & 659.7~ & ~~0.38(2)$^{\rm a}$ \\
& others  & & $\ge$ 3000~~ & ~0.011(2)$^{\rm b}$ \\
\hline
\end{tabular}
$\rm{^a}$ from reference \cite{NIGGLI1987} and $\rm{^b}$  this work
\cite{sdthesis}.
\end{table}

Two different repumping schemes were investigated, where repumping
was implemented via low lying states to minimize the transfer of
atomic population into additional states. The first scheme uses the
6s6p~$^{1}$P$_{1}$ level as an intermediate state. This constitutes
a closed five-level manifold (6s$^2$~$^1$S$_0$, 6s6p~$^1$P$_1$,
6s5d~$^1$D$_2$, 6s5d~$^3$D$_2$ and 6s5d~$^3$D$_1$) involving
infrared transitions at the wavelengths $\lambda_{\rm{ir1}}$,
$\lambda_{\rm{ir2}}$ and $\lambda_{\rm{ir3}}$ (Tab.
\ref{bawavelengths}). The common excited state for the cooling
transition and the repumping transitions leads to multiple coherent
Raman resonances \cite{DAMMALAPATI2006}. In the limit of high
intensities the population in all five states is equal. This reduces
the maximum optical force by a factor of 2/5 compared to an ideal
closed two level system. In atoms with nuclear spin different states
of the hyperfine manifold can be employed for cooling and repumping.
The second scheme alternatively repumps the 6s5d~$^3$D$_{1}$-state
through the 5d6p~$^3$D$_1^\circ$ state using light at wavelength
$\lambda_3$. This strong transition exhibits only a 2.0(4)$\%$ leak
to further states. The main contribution arises from the
5d$^2$~$^3$F$_2$-state, which cascades to 94(3)$\%$ down to one of
the states of the cooling manifold according to a recent calculation
\cite{flambaumlevelcalc}.

In the experiment, an Ba atomic beam is produced from an
isotopically enriched sample of $^{138}$BaCO$_3$ which is mixed with
Zr powder as a reducing agent in a resistively heated oven at
780(40)~K temperature (Fig. \ref{coolsetup}). The beam enters a
straight section, where it is overlapped with counter-propagating
deceleration laser beams at wavelengths $\lambda_1$,
$\lambda_{\rm{ir1}}$ and $\lambda_{\rm{ir2}}$ from one dye-laser and
two diode lasers. The laser beams are spatially overlapped and
focussed into the 1~mm diameter oven orifice. The divergence of the
laser beams is typically 5~mrad, their diameter 600~mm downstream of
the oven is 3~mm and the typical laser powers are 12 mW at
wavelength $\lambda_1$, 5mW at $\lambda_{\rm{ir1}}$ and 35~mW at
$\lambda_{\rm{ir2}}$. The detunings from the resonances are
-290(2)~MHz, -60(10)~MHz and -90(10)~MHz, respectively. These
parameters permit slowing of atoms with velocities up to 150~m/s.
Additional repumping from the 6s5d~$^3$D$_1$ state allows for even
larger velocity changes.

At the end of the slowing region three mutually orthogonal beams of
12~mm diameter of up to 15~mW power at wavelength $\lambda_1$ are
retro-reflected into themselves. The required circular polarizations
for a MOT are produced by a set of $\lambda$/4-plates. The six beams
are overlapped in the minimum of a magnetic field produced by a pair
of coils in anti-Helmholtz configuration, which creates a field
gradient along the axis between the two coils of up to 36~G/cm. The
frequency detuning from resonance $\delta_{\rm{trap}}$ of the
trapping light can be adjusted between -200~MHz and 50~MHz by
acousto optical modulators (AOM's). Three custom-made fiber lasers
(Koheras) at the wavelengths $\lambda_{\rm{ir1}}$ (5~mW),
$\lambda_{\rm{ir2}}$ (25~mW) and $\lambda_{\rm{ir3}}$ (60mW) of
typically 5~mm diameter are overlapped with the central region of
the trap. The detuning from resonance is small compared to the
linewidth of 18~MHz of these transitions to achieve optimal
repumping for the trapped atoms. For the second repumping scheme a
laser beam at wavelength $\lambda_3$ of 10~mm diameter and 5~mW
light power is co-propagating with the MOT beams.

The fluorescence from the central region of the trap is collected by
a 60~mm focal length plano-convex lens. An aperture of 2.0(5)mm
diameter in the image plane allows to select the field of view.
Fluorescence at the wavelengths $\lambda_1$ and $\lambda_{\rm{B}}$
are detected by two photomultiplier tubes (PMT's) equipped with
interference filters of 10~nm spectral bandwidth.
\begin{figure}[tb]
\center
\includegraphics[width = 82mm]{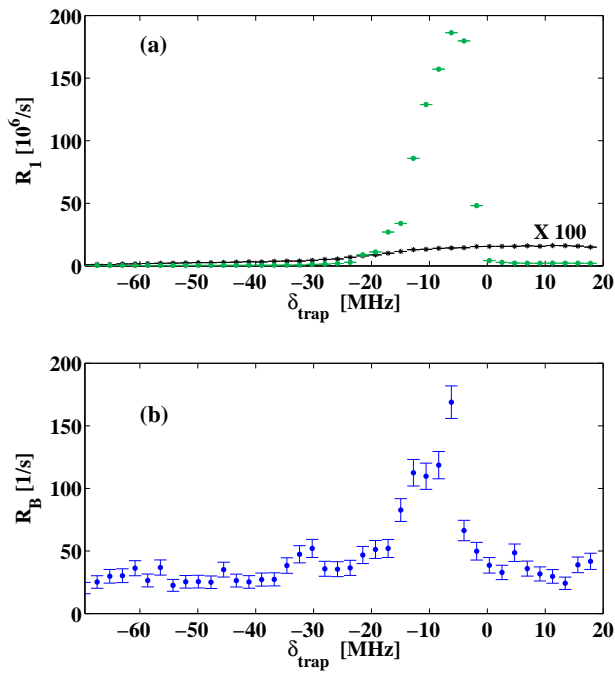}\\
\caption{Signals from trapped atoms as a function of detuning of the
trapping laser light at wavelength $\lambda_1$.(a) Fluorescence at
wavelength $\lambda_1$. The black points is the Doppler-free
fluorescence signal arising from the MOT laser beam which is
orthogonal to the atomic beam.(b) Fluorescence at wavelength
$\lambda_{\rm{B}}$ detected simultaneously. The trap lifetime was
$\tau_{\rm{MOT}}$=0.15(4)~s.} \label{lambda1}
\end{figure}

The detected fluorescence increases for small negative frequency
detunings $\delta_{\rm{trap}}$ of the trapping laser beams at
wavelength $\lambda_1$ (Fig. \ref{lambda1} (a)). The vertical
MOT-beam, which is orthogonal to the atomic beam, produces a Zeeman
broadened fluorescence signal which can be used to estimate the flux
of atoms in the atomic beam. A comparison of the signal rates for
these two conditions yields the fraction of the atomic beam, which
is captured in the MOT. The scattering rate from the MOT laser beams
is about the same for trapped atoms and from the beam. With this
assumption the collection efficiency is
\begin{equation}
\epsilon=\frac{\rm{R}_{1}}{\rm{R_{beam}}}\cdot
\frac{\Delta\rm{t}}{\tau_{\rm{MOT}}},
\end{equation}
where R$_{1}$ and R$_{\rm{beam}}$ are the fluorescence rates from
trapped atoms and from the Doppler-free signal of the beam and
$\Delta$t is the average time of flight of thermal atoms through the
light collection region. An efficiency of
$\epsilon$=0.4(1)$\cdot$10$^{-2}$ is determined. A trap population
$\rm{N_{MOT}}$ of up to $10^6$ atoms and a lowest temperature for
the trapped cloud of 5.4(7)~mK were achieved.

The fraction of atoms in the metastable D-states $\rm{\rho_D}$ can
be determined from the fluorescence rates $\rm{R_B}$ at wavelength
$\rm{\lambda_B}$ produced by the repumping the 6s5d~$^3$D$_1$ state
via the 5d6p~$^3$D$_1^{\rm o}$-state (Fig. \ref{lambda1} (b)). The
detected rate R$_{\rm{B}}$ is
\begin{equation}
\rm{R_{B}=\epsilon_{\rm{B}} \cdot \rm{B_B} \cdot \rm{B_{ir3}} \cdot
\gamma_1 \cdot \rm{N_{MOT}}},
\end{equation}
where $\rm{\epsilon_B}$ is the detection efficiency for a photon at
wavelength $\rm{\lambda_B}$, $\rm{B_B}$=2.2(2)$\%$ is the decay
branching fraction of the 5d6p~$^3$D$_1^{\rm o}$-state to the ground
state, $\rm{B_{ir3}}$=4.2(4)$\cdot$10$^{-5}$ is the decay branching
fraction of the 6s6p~$^1$P$_1$-state to the 6s5d~$^3$D$_1$-state and
$\gamma_1$ the scattering rate at wavelength $\lambda_1$. Similarly
the rate $\rm{R_1}$ at wavelength $\lambda_1$ is
\begin{equation}
\rm{R_{1}}=\epsilon_1 \cdot \gamma_1 \cdot \rm{N_{MOT}} \cdot
(1-\rho_D).
\end{equation}
The ratio of  the detection efficiencies was measured to
$\epsilon_{\rm{B}}$:$\epsilon_1$=1:0.8(1). The observed rather large
fraction of $\rho_D$=0.5(1) is due to the strong coherent Raman
transitions in the cooling scheme.

\begin{figure}[b]
\center
\includegraphics[width = 78mm, angle = 0]{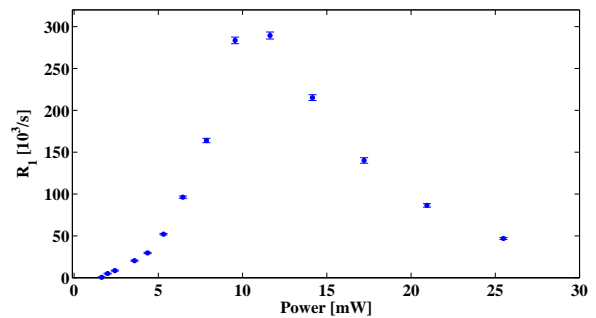}\\
\caption{Dependence of the number of trapped atoms $\rm{N_{MOT}}$ on
the laser power of the deceleration beam at wavelength $\lambda_1$.
The decrease is due to stopping of atoms before they have reached
the trapping region.} \label{slowingpower}
\end{figure}
The number of trapped atoms depends on the intensity of the slowing
beam at wavelength $\lambda_1$ (Fig. \ref{slowingpower}). The
detuning of the slowing laser beam was $\delta_{\rm{s}}$ = -260~MHz,
corresponding to a velocity class of 145~m/s. The number of trapped
atoms is proportional to the loading rate. The loading rate
increased up to a cooling beam power of about 11~mW which
corresponds to 2.7 saturation intensities. A further increase of the
cooling beam power does not increase the flux into the MOT. This is
caused by stopping of atoms before they have reached the trapping
region. Thus, the velocity change in l=600~mm length is larger than
150~m/s. The average deceleration exceeds 1.7$\cdot 10^{4}$~m/s$^2$.
The flux at low velocities can be significantly improved by a weak
laser beam at wavelength $\lambda_1$ co-propagating with the atomic
beam, which would define a finite end velocity \cite{whitelight}.

\begin{figure}[tb]
\center
\includegraphics[width = 78mm, angle = 0]{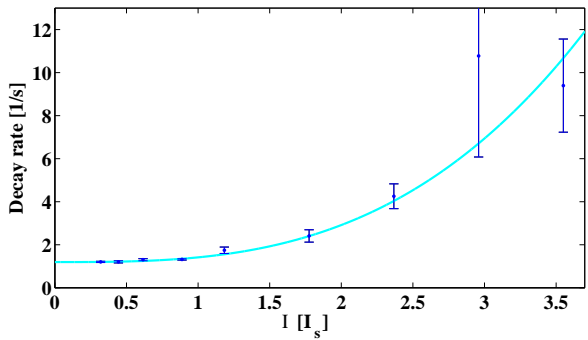}\\
\caption{Decay rate from trapped sample as a function of MOT laser
intensity. The solid line is
1/$\tau_{\rm{MOT}}=1/\tau_{\rm{MOT}}^0+\beta\cdot \rm{I}^3$, where
$1/\tau_{\rm{MOT}}^0$ is the decay rate at intensity I=0, and
$\beta$ is the rate constant.} \label{decayrate}
\end{figure}

The lifetime $\tau_{\rm{MOT}}$ of the trapped sample depends
strongly on the intensity of the trapping laser beams at wavelength
$\lambda_1$ (Fig. \ref{decayrate}). A 3rd-order process is observed
for the losses as a function of laser intensity I with a rate
constant of $\beta$=0.20(3)~s$^{-1}$~I$_{\rm{s}}^{-3}$, where
I$_{\rm{s}}$=14.1 mW/cm$^2$ is the saturation intensity. This could
be explained by three-photon ionization. The MOT lifetime
$\tau_{\rm{MOT}}^0$ of up to 1.5~s depends on the intensity in the
repumping laser beams. The overlap of all laser beams in the trap
region is crucial. Atoms in the metastable states can escape from
the trap, since they do not experience any trapping force. Similar
effects have been observed in Ca \cite{alcatraz}.

We have realized laser cooling of Ba in a closed five-level
subsystem and a six-level system with a small leak. These are the
minimal subsets of levels for laser cooling of Ba in the ground
state. Efficient deceleration of the atomic beam is achieved with
counter-propagating lasers at high intensities. Further improvements
can be expected from frequency broadening of the deceleration
lasers, e.g. with electro-optical modulation. This would enlarge the
velocity acceptance of the deceleration and a larger fraction of the
atomic beam velocity distribution can be stopped. As a note, a
Zeeman slower is not applicable in such multilevel atomic laser
cooling systems, for which repumping during slowing is required,
because the changing Doppler shifts cannot be compensated by single
magnetic field for all transitions at the same time. In particular,
some of the repumping transitions in Ba have a negative value for
the g-factor $\rm{g_F}$.

The demonstrated laser cooling with a complex cooling cycle appears
particularly well suited for efficient collection of rare isotopes
in a MOT with similar level schemes, i.e. Ra isotopes. The strong
motivation for trapping Ra arises because such samples allow for
novel precision measurements within the Standard Model in particle
physics and searches for physics beyond it such as they are underway
at the KVI TRI$\mu$P facility \cite{traykov}. Atomic structure
calculations for heavy alkaline earth atoms to evaluate their
sensitivity to new physics are performed by several groups
\cite{atomicstructure}. Recent computations show that Ra offers an
enhancement of about 500 times due to nuclear effects
\cite{engeletc} and 40000 times due to the unique atomic level
structure for nucleon or electron EDM's \cite{flambaum1999}. In
addition, APV induced effects are 100 times larger for the weak
charge and 10$^3$ times larger for the nuclear anapole moment than
in other systems \cite{flambaum2000}. All Ra isotopes with nuclear
spin I$\ne$0 have short lifetimes (e.g. $^{225}$Ra, $\tau$~=~14.8 d)
and are only available in small quantities and require experiments
in the proximity of an isotope production facility. Sensitive
experimental searches for EDM's, which would establish CP violation
without strangeness, and measurements of APV in Ra require the
developed trapping techniques.

This work was supported by the \textit{Nederlandse Organisatie voor
Wetenschappelijk Onderzoek (NWO)} by an NWO-VIDI grant (No.
639.052.205) and the \textit{Stichting voor Fundamenteel Onderzoek
der Materie (FOM)} under programme 48 (TRI$\mu$P).

\end{document}